\documentclass[prc,
showpacs,nofootinbib,
superscriptaddress,
twocolumn
]{revtex4-1}
\usepackage{amsmath,graphicx}
\usepackage{braket,hyperref}

\begin{document}

\title{Large-Scale Calculations of the Double-Beta Decay of $^{76}$Ge,
$^{130}$Te, $^{136}$Xe, and $^{150}$Nd in the Deformed Self-Consistent Skyrme
Quasiparticle Random-Phase Approximation} 
\date{\today}
\author{M. T. Mustonen} \email{mika.t.mustonen@unc.edu}
\affiliation{Department of Physics and Astronomy, CB 3255, University of North Carolina, Chapel Hill, NC 27599-3255}
\affiliation{Department of Physics, Central Michigan University, Mount Pleasant, MI 48859}
\author{J. Engel} \email{engelj@physics.unc.edu}
\affiliation{Department of Physics and Astronomy, CB 3255, University of North
Carolina, Chapel Hill, NC 27599-3255} 

\begin{abstract}
We use the axially-deformed Skyrme Quasiparticle Random Phase Approximation
(QRPA) together with the SkM$^*$ energy-density functional, both as
originally presented and with the time-odd part adjusted to reproduce the
Gamow-Teller resonance energy in $^{208}$Pb, to calculate the matrix elements
governing the neutrinoless double-beta decay of $^{76}$Ge, $^{130}$Te,
$^{136}$Xe, and $^{150}$Nd.  Our matrix elements in $^{130}$Te and $^{136}$Xe
are significantly smaller than those of previous QRPA calculations, primarily
because of the difference in pairing or deformation between the initial and
final nuclei.  In $^{76}$Ge and $^{150}$Nd our results are similar to those of
less computationally intensive QRPA calculations.  We suspect the $^{76}$Ge
result, however, because we are forced to use a spherical ground-state, even
though the HFB indicates a deformed minimum.  
\end{abstract}

\maketitle

\section{Introduction}

Neutrinoless ($0\nu\beta\beta$) double-beta decay can occur if neutrinos are
Majorana particles, at a rate that depends on a weighted average of neutrino
masses (see Refs.\ \cite{avi08,Ver12} for reviews).  The experimental search
$0\nu\beta\beta$ is approaching sensitivity to neutrino masses below 100 eV.
Extracting a mass from the results, however, or setting a reliable upper limit,
will require accurate values of the nuclear matrix elements governing the
decay, matrix elements that cannot be measured and must therefore be
calculated.  A number of theorists have attempted the calculations, applying
several distinct methods.  Among the most popular is the proton-neutron
quasiparticle random phase approximation (QRPA).  

The QRPA can be carried out at various levels of sophistication.  So far, with
only a few exceptions \cite{Sim04,Fan11}, the mean-fields on which the QRPA is
based have been spherical by fiat.  In addition, they have never been
consistent with the residual QRPA interaction, nor have all the nucleons ever
been active degrees of freedom, free from confinement in an artificially inert
core.  Finally, even the active nucleons have been forced to occupy a few
harmonic-oscillator shells and thus have never tasted the continuum.  Here we
overcome all these limitations, allowing axially symmetric deformation, using a
modern and well-tested Skyrme functional for both the Hartree-Fock-Bogoliubov
(HFB) mean-field calculation and the QRPA that is based on it, keeping all
the nucleons active, and placing the nucleus inside a large cylindrical box,
so that discretized versions of continuum states up to high energy are
available.  

Deformed Skyrme-QRPA calculations of this type have been applied extensively in
recent years to nuclear vibrations (see e.g.,
\cite{Ter10,Ter11,Yos11,Per11,Yos08}) and will soon be applied to single-beta
decay \cite{Tom-WIP}.  Our implementation, described in detail below, is via a
B-spline-based HFB code with the above-mentioned cylindrical-box boundary
conditions followed by the construction and diagonalization of the QRPA
Hamiltonian matrix in the basis of canonical two-quasiparticle states.  The
calculations consume enough CPU hours to require a supercomputer, and so we
restrict ourselves here to four isotopes --- $^{76}$Ge, $^{130}$Te, $^{136}$Xe
and $^{150}$Nd --- used in the some of the most promising of current or
proposed experiments.  The deformation and pairing in the initial and final
nuclei are often quite different and matrix elements can be suppressed as a
result \cite{Sim04}; our numbers depend crucially on the overlap of
intermediate-nucleus states created by exciting the initial ground state with
those created by exciting the final ground state.  The QRPA supplies only
transition amplitudes and so must be extended to obtain the overlap.  Here we
will apply a prescription like that in Ref.\ \cite{Sim04}, while noting that a
well justified and tractable expression is still lacking. 

This article is organized as follows: Section \ref{sec:theory} contains a brief
overview of the matrix elements governing double-beta decay and of the Skyrme
QRPA.  Section \ref{sec:comp} describes the details of our computational
implementation and Sec. \ref{sec:results} presents our results.  Section
\ref{sec:conc} is a conclusion.

\section{Double-beta decay and the QRPA}\label{sec:theory}

\subsection{Decay operators}
The lifetime for $0\nu\beta\beta$ decay, if there are no heavy particles
mediating the decay, is 
\begin{equation}
[T_{1/2}^{0\nu}]^{-1} = G'^{0\nu} \braket{m_\nu}^2 | M'^{0\nu} |^2 \,,
\end{equation}
where $\braket{m_\nu}^2$ is a weighted average of three neutrino masses,
$G'^{0\nu}$ is a phase space factor (recently recomputed in Ref.\
\cite{Kot12}), and $M'^{0\nu}$ is a nuclear matrix element\footnote{This matrix
element differs from the unprimed $M^{0\nu}$ used elsewhere by a factor of
$g_A^2/1.25^2$. The two are equivalent when $g_A$ is taken to be 1.25, but
differ when it is modified.  (Actually, $g_A$ is closer to 1.27 than 1.25, but
we follow tradition here.)  The convention we use puts all the $g_A$ dependence
in the matrix element and none in the phase-space factor.}.  Although the
matrix element contains intermediate states and an energy denominator, it can
to good approximation \cite{pan90} be represented by one involving only the
initial and final ground states.  In this ``closure'' approximation and
neglecting the small tensor term, one can write the matrix element as
\begin{align}
\label{eq:me}
M'^{0\nu} & =  \frac{2R}{\pi (1.25)^2} \int_0^\infty \!\!\! q \, dq  \\ 
 &\times \bra{f} \sum_{a,b}\frac{j_0(qr_{ab})\left[ h_F(q)+   
 h_{GT}(q) \vec{\sigma}_a \cdot \vec{\sigma}_b \right]}
 {q+\overline{E}-(E_i+E_f)/2} \tau^+_a \tau^+_b \nonumber \ket{i} \,,
\end{align}
where the factor 1.25 is inserted by convention, $r_{ab}=|\vec{r}_a-\vec{r}_b|$
is the distance between nucleons $a$ and $b$, $j_0$ is the usual spherical
Bessel function, $\bar{E}$ is an average excitation energy to which the matrix
element is insensitive (and for which we use the value 10 MeV), and the nuclear
radius $R\equiv 1.2 A^{1/3}$ fm is inserted with a compensating factor in the
phase-space function to make the matrix element dimensionless.  The ``form
factors'' $h_F$ and $h_{GT}$ are given by
\begin{align}
\label{eq:form-facs}
h_F(q) &\equiv -g_V^2(q^2)\\
h_{GT}(q)&\equiv g_A^2(q^2) -\frac{g_A(q^2) g_P(q^2) q^2}{3m_p} +
\frac{g_P^2(q^2)q^4}{12m_p^2} \nonumber \\
&+ \frac{g_M^2(q^2)q^2}{6m_p^2} \,, \nonumber 
\end{align}
with 
\begin{align}
\label{eq:dipoles}
g_V(q^2)&=\frac{1}{\left(1+q^2/(0.71 \textrm{ GeV}^2)\right)^2} \\
g_A(q^2)&=\frac{1.27}{\left(1+q^2/(1.09 \textrm{ GeV}^2)\right)^2} \nonumber\\
g_P(q^2)&=\frac{2m_p g_A(q^2)}{q^2+m_\pi^2}  \qquad g_M(q^2) = 3.70
g_V(q^2) \,.  \nonumber
\end{align}
Here $m_p$ and $m_\pi$ are the proton and pion masses.

The two-neutrino double-beta decay ($2\nu\beta\beta$) rate, which we will use
to fit parameters for our $0\nu\beta\beta$ calculation, can be written as
\begin{equation}
[T_{1/2}^{2\nu}]^{-1} = G^{2\nu} | M^{2\nu} |^2.
\end{equation}
where $G^{2\nu}$ is another phase-space factor (also recomputed in
Ref.~\cite{Kot12}) and $M^{2\nu}$ is a matrix element.  The closure
approximation is not good for two-neutrino decay, and the matrix element must
contain intermediate states explicitly:
\begin{equation}
\label{eq:two-nu}
M^{2\nu}  \approx \sum_n \frac{\bra{f}\sum_a \vec{\sigma}_a \tau^+_a
\ket{n} \bra{n}\sum_b
\vec{\sigma}_b \tau^+_b \ket{i}}{E_n-(M_i+M_f)/2} \,,
\end{equation}
where $n$ labels states in the intermediate nucleus with energy $E_n$, $M_i$
and $M_f$ are the masses of initial and final nuclei, and the effects we've
neglected --- forbidden currents, the Fermi matrix element, etc.\ --- are
small.  

Recent study \cite{sim09,eng09} has shown that realistic short-range
correlations have only a small effect on the double-beta matrix elements.
Including them here, even approximately, would complicate our computational
procedure considerably and so we omit them altogether.

\subsection{Deformed Charge-changing QRPA}

The self-consistent axially-symmetric Skyrme-HFB-QRPA method for like-particle
excitations, on which our code is based, is described thoroughly in Refs.\
\cite{Ter05}, \cite{Ter10}, and \cite{Ter11}.  We modify the code discussed
there in a rather straightforward way --- changing the basis of
like-two-quasiparticle states to a basis of one-quasiproton-one-quasineutron
states, removing the Coulomb interaction, and keeping only the relevant parts
of the Skyrme functional --- to work with charge-changing modes rather than
like-particle modes.  

We adopt the Skyrme functional (or effective interaction) SkM$^*$ \cite{Bar82};
that functional has been shown to describe nuclear deformation well and
reproduces low-lying quadrupole vibrations in rare-earth nuclei noticeably
better than the comparably popular functional SLy4 \cite{Ter10}.  We modify the
time-odd particle-hole part of the functional as in \cite{Ben02}, which
discussed charge-changing transitions, by setting the parameters (defined in
that reference) $C_1^T=0$, $C_1^{\nabla s}=0$, and $C_1^s[0] =
C_1^s[\rho_\text{nm}] = 100$ Mev fm$^3$ ($\rho_\text{nm}$ is nuclear-matter
density).  With these modifications, the functional reproduces \cite{Tom-WIP}
the location of the Gamow-Teller resonance and the fraction of observable
strength in the resonance.  We will report results with and without the
modifications to show their effect.
 
For the particle-particle part of the functional we use a simple volume
(zero-range) pairing interaction, the strength of which we adjust separately in
the isoscalar channel ($T=0$) and in each of the three isovector ($T=1$ with
$T_z=-1$, 0, and 1) channels.  We describe the adjustment in more detail in the
next section.

Evaluating the $0\nu\beta\beta$ matrix elements requires a multipole
decomposition of $M'^{0\nu}$ suitable for cylindrical geometry.  The details of
that appear in the Appendix.

\section{Computational implementation}\label{sec:comp}

Only recently have fully self-consistent deformed Skyrme-QRPA calculations
entered the scene.  The combination of methods we use here requires many
thousands of CPU hours.  Our methodology will at some point be obsolete because
of the development of much faster Finite Amplitude \cite{Avo11} and iterative
Arnoldi \cite{Toi10} approaches, which use mean-field codes with
time-independent constraints to solve the QRPA equations.  Our method, by
contrast, involves the explicit construction and diagonalization of the QRPA
Hamiltonian matrix in a basis of two-canonical-quasiparticle states.  These
states are obtained from the HFB calculation mentioned previously. 

To solve the initial HFB equations we use the Vanderbilt deformed HFB code
\cite{Ter03}, which represents wave functions in a basis of B-splines.  Our
cylindrical box has dimensions $r_\mathrm{max} = z_\mathrm{max} = 20$ fm, about
three times as large as the radius of the heaviest nucleus studied here and a
number found suitable in Ref.~\cite{Ter03}.  Our mesh spacing is 0.7 fm and the
energy cutoff of the HFB solutions is 60 MeV.  We do not restrict the
deformation (except to be axially symmetric) but rather allow the mean field to
evolve freely to the nearest local binding-energy minimum.  Using a range of
quadrupole deformation parameters $\beta_2$ as initial guesses, we find one or
more local minima and select the most bound solution as the mean field on which
we base the QRPA.  In $^{76}$Ge, however, we do not use the most bound
solution; we discuss the reasons for this exception in the next section.  To
obtain the strength of the proton-proton and neutron-neutron ($T=1, T_z=\pm 1$)
pairing interaction we match the HFB pairing gaps with the experimental pairing
gaps obtained from a three-point interpolation formula, with separation
energies from the ENSDF \cite{ensdf} database.  

The computational requirements for running our charge-changing QRPA code are
significantly less than those for the like-particle code on which it is based
because a) the proton-neutron two-quasiparticle basis is only about half the
size of the like-two-quasiparticle basis, and b) the removal of the Coulomb
interaction relieves us of a large computational burden.  For a given
multipole, our charge-changing code typically runs much faster than the
like-particle code.  That speedup, however, still leaves us with runs that
consume many thousands of CPU hours per multipole in each nucleus.

We cannot include all one-quasiproton-one-quasineutron states in our QRPA basis
and so truncate the same way as in Ref.\ \cite{Ter10}.  The truncation is
controlled by two parameters $v^\mathrm{pp}_\mathrm{cut}$ and
$v^\mathrm{ph}_\mathrm{cut}$.  The first allows us to remove two-quasiparticle
states with almost completely two-particle or two-hole nature (i.e.\ states
that primarily lie in the $A \pm 2$ neighbors of the reference nucleus instead
of the in intermediate double-beta nucleus), and the second lets us cut out
states in which one of the particles is far below the Fermi surface and the
other far above it.  Such excitations have very high energy and do not mix
significantly with lower-energy states.  In practice 15,000 two-quasiparticle
states for the lowest multipoles, out of a total of about half a million, are
enough to approximate the exact answer very well, making the construction and
diagonalization of the QRPA matrix tractable on a supercomputer.

After diagonalizing the QRPA Hamiltonian, we need to determine the
double-beta-decay matrix elements.  For $0\nu\beta\beta$ decay, the matrix
element can be written as 
\begin{align}
\label{monu-detail}
M'^{0\nu} =& \frac{2R}{(1.25)^2\pi} 
	\sum_{pn} \langle 0^+_f | c_{-p}^\dag c_n | N \rangle
	\sum_{NN'} \langle N | N' \rangle \\
	&\times \sum_{p'n'} \langle N' | c_{p'}^\dag c_{-n'} | 0^+_i \rangle
	\left( K^{F}_{pn,p'n'} + K^{GT}_{pn,p'n'} \right) \,, \nonumber
\end{align} 
where $c^\dag_k$ are particle-creation operators, the indices with $p$ refer to
protons and those with $n$ to neutrons, each index stands for the set of
quantum numbers $p=\{j^z_p, \pi_p, k_p \}$ (angular-momentum along on the
intrinsic axis, parity, and an additional enumerating index), a minus sign in
front of an index means that the sign of the $j^z$ quantum number is reversed,
and 
\begin{align}
\label{eq:monusp}
K^{F}_{pn,p'n'} &= 
 \int_0^\infty \!\!\! q \, dq   
 \bra{pp'} \frac{j_0(qr_{12}) h_{F}(q)} 
 {q+\overline{E}-(E_i+E_f)/2} \tau^+_1 \tau^+_2 \ket{nn'} \\
K^{GT}_{pn,p'n'} &= 
 \int_0^\infty \!\!\! q \, dq   
 \bra{pp'} \frac{j_0(qr_{12}) h_{GT}(q) 
 \vec{\sigma}_1 \cdot \vec{\sigma}_2 } 
 {q+\overline{E}-(E_i+E_f)/2} \tau^+_1 \tau^+_2 \ket{nn'} \,. \nonumber
\end{align}
The two-particle states in Eq.\ \eqref{eq:monusp} are antisymmetrized.  We use
a multipole expansion, detailed in the Appendix, to evaluate the two-body
matrix elements in Eq.\ \eqref{eq:monusp} with B-spline integration.  The
coding for the two-neutrino two-body matrix elements, which we use to evaluate
the matrix element in Eq.\ \eqref{eq:two-nu}, requires no Bessel function
expansion.

\begin{figure}[t]
\centering
\includegraphics[scale=1]{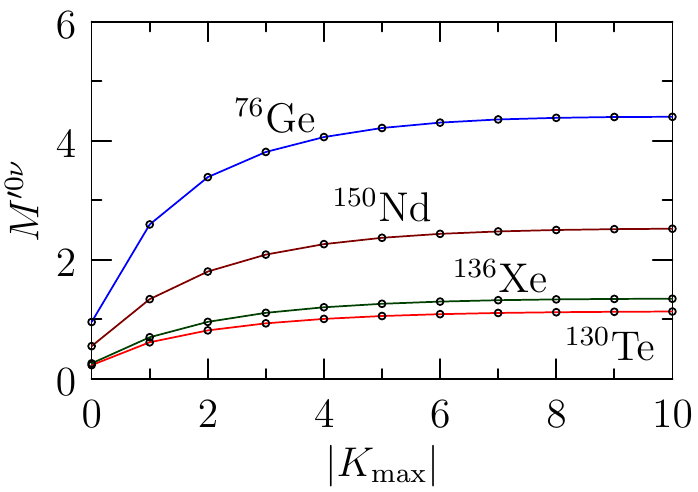}
\caption{(Color online.) The cumulative $^{76}$Ge $0\nu\beta\beta$ matrix
element (using SkM$^*$ and $g_A=1.0$) as the number of intermediate-state
multipoles $K^\pi$ is increased.  Convergence is reached by $|K|=10$.  Both
positive and negative parities are included, as are both Fermi and Gamow-Teller
contributions.} 
\label{fig:Kconvergence}
\end{figure}

Two-neutrino decay is simpler for another reason as well; only states with
angular momentum and parity $J^\pi = 1^+$ contribute to the matrix element.  In
our deformed calculation we follow the usual procedure of representing
laboratory states in a rigid-rotor approximation as combinations of a) Wigner
functions $D^J_{MK}$ and $D^J_{M-K}$ of Euler angles, and b) an intrinsic QRPA
state with a well-defined projection $K$ along the symmetry axis of the angular
momentum $\vec{J}$.  $M^{2\nu}$ thus gets contributions only from states with
$|K| \le 1$.  In neutrinoless decay, on the other hand, states with any $K^\pi$
contribute.  The contributions get progressively smaller as $K$ gets larger.
Including state with $|K| \le 10$ is enough to approximate the matrix element
accurately, as Fig.~\ref{fig:Kconvergence} shows.  

One interesting feature of Eq.\ \eqref{monu-detail} is the presence of the
overlap $\braket{N|N'}$.  The QRPA is a small-amplitude approximation and
although it provides transition densities from a ground state to excited
states, it can't, without extension, provide excited state wave functions.  The
excited states $\ket{N}$ and $\ket{N'}$ are based on different quasiparticle
vacua and the quasiboson approximation that is inherent in the QRPA erases the
information necessary to relate the two vacua.  Two expressions for the overlap
have been given in the past few years: one, from Ref.\ \cite{Sim04}, neglects
``scattering terms'' even though they cannot be shown to be small and the
other, laid out in Ref.\ \cite{Ter12}, uses the form of the boson vacuum but
replaces the bosons with the fermion pairs from which they stem.
Unfortunately, this last idea leads to expressions that can only be evaluated
perturbatively; these become unwieldy after the lowest couple of orders in the
expansion, the convergence of which may not be fast.  Here we simply evaluate
the overlap in the quasi-Tamm-Dancoff approximation (neglecting the QRPA ``Y''
amplitudes); in this limit of the QRPA, excited states are well-defined
two-quasiparticle excitations of HFB vacua and require no bosonization.  The
results are not very different from those obtained in the scheme proposed in
Ref.\ \cite{Sim04}.  We provide more details in the Appendix.  

As is typical in QRPA calculations, we use the measured values of two-neutrino
decay rates to fit proton-neutron pairing strengths.  Following a suggestion in
Ref.\ \cite{Rod11}, we adjust the isovector ($T=1, T_z=0$) strength so that the
Fermi $2\nu\beta\beta$ matrix element vanishes, as it should (almost) because
the ground state of the final nucleus has a different isospin than the
double-isobar-analog state of the initial nucleus.  If instead we fix the
proton-neutron isovector pairing strength at the average of the proton-proton
and neutron-neutron pairing strengths, we find a nearly identical result.  The
isoscalar pairing strength, which we call $V_0$ here, is the parameter
typically called $g_{pp}$ in other QRPA calculations.  We adjust it so as to
reproduce the experimental two-neutrino matrix element, with both an unquenched
($g_A=1.25$, see the footnote) and quenched ($g_A=1.0$) axial-vector coupling
constant.  We then use the resulting pairing strengths in computing the
$0\nu\beta\beta$ matrix elements, once for each value of $g_A$.  For $^{130}$Te
and $^{136}$Xe, we compute the neutrinoless double-beta-decay matrix element
with the unmodified SkM$^*$ over a range of isoscalar pairing values $V_0$ to
assess its sensitivity to the fit.

\begin{table}[!t]
	\begin{tabular}{l c c c c c}
	\hline \hline
	& & \multicolumn{2}{c}{Ref.~\cite{Alv04}} & \multicolumn{2}{c}{Exp.} \\
	\cline{3-4}\cline{5-6} & this work & Sk3 & SG2 & Ref.~\cite{Lal99} &
	Ref.~\cite{Ram01} \\
	\hline
	$^{76}$Ge & 0.184\footnote{-0.025 used} & 0.161 & 0.157 & 0.095(30) & 0.2623(9) \\
	$^{76}$Se & -0.018 & -0.181 & -0.191 & 0.163(33) & 0.3090(37) \\
	$^{130}$Te & 0.01 & -0.076 & -0.039 & 0.035(23) & 0.1184(14) \\
	$^{130}$Xe & 0.13 & 0.108 & 0.161 & - & 0.1837(49) \\
	$^{136}$Xe & 0.004 & 0.001 & 0.016 & - & 0.122(10) \\
	$^{136}$Ba & -0.021 & 0.009 & 0.070 & - & 0.1258(12) \\
	$^{150}$Nd & 0.27 & 0.266 & 0.271 & 0.367(86) & 0.2853(21) \\
	$^{150}$Sm & 0.22 & 0.207 & 0.203 & 0.230(30) & 0.1931(21) \\
	\hline\hline
	\end{tabular}
	\caption{The quadrupole deformations $\beta_2$ of the initial and final
	nuclei in our work, compared with the values obtained in \cite{Alv04}
	and experimental values from \cite{Lal99,Ram01}}. 
	\label{tab:deformation}
\end{table}

\section{Results and discussion}\label{sec:results}

We start by comparing the quadrupole deformation parameters $\beta_2$ obtained
from our HFB calculation to other theoretical and experimental values in Tab.\
\ref{tab:deformation}.  With the exception of $^{76}$Se, where our Skyrme-HFB
computation fails to converge to a prolate solution, our quadrupole
deformations are similar to those obtained using Sk3 and SG2 Skyrme
interactions in Ref.~\cite{Alv04}.  The failure to converge is most likely due
to a very flat bottom of the binding energy curve with respect to deformation
in $^{76}$Se. 

In $^{76}$Ge, the minimum energy occurs at a prolate deformation of
$\beta_2=0.18$.  This deformation is so different from that of $^{76}$Se,
however, that our predicted two-neutrino matrix element is smaller than the
measured value no matter what we use for $g_A$ or $V_0$.  We therefore choose
to use the local near-spherical minimum ($\beta_2=-0.025$) for $^{76}$Ge
instead.  As we shall see, this gives us a result that is not too different
from other QRPA numbers, including those of Ref.\ \cite{Fan11}, which presents
both spherical-spherical and prolate-prolate calculations.  It also indicates,
however, that the QRPA is inadequate in this system.  The soft surfaces with
multiple minima require a formulation that mixes mean fields, e.g.\ the
generator-coordinate method (often referred to as energy-density functional
(EDF) theory) of Ref.\ \cite{Rod10}, or an extension thereof.  

In the daughter nucleus $^{130}$Xe we get a prolate solution, making ours the
first QRPA calculation to take the deformation into account in the decay of
$^{130}$Te.  The study in Ref.~\cite{Rob08}, using HFB with the Gogny
interaction, finds a second minimum with oblate deformation and a barrier of
only 1 MeV or so separating the two minima.  As in $^{76}$Se, therefore, the
use of a single mean-field in the construction of the $^{130}$Xe ground state
is somewhat suspect. 

\begin{figure}[t]
\centering
\includegraphics[width=\columnwidth]{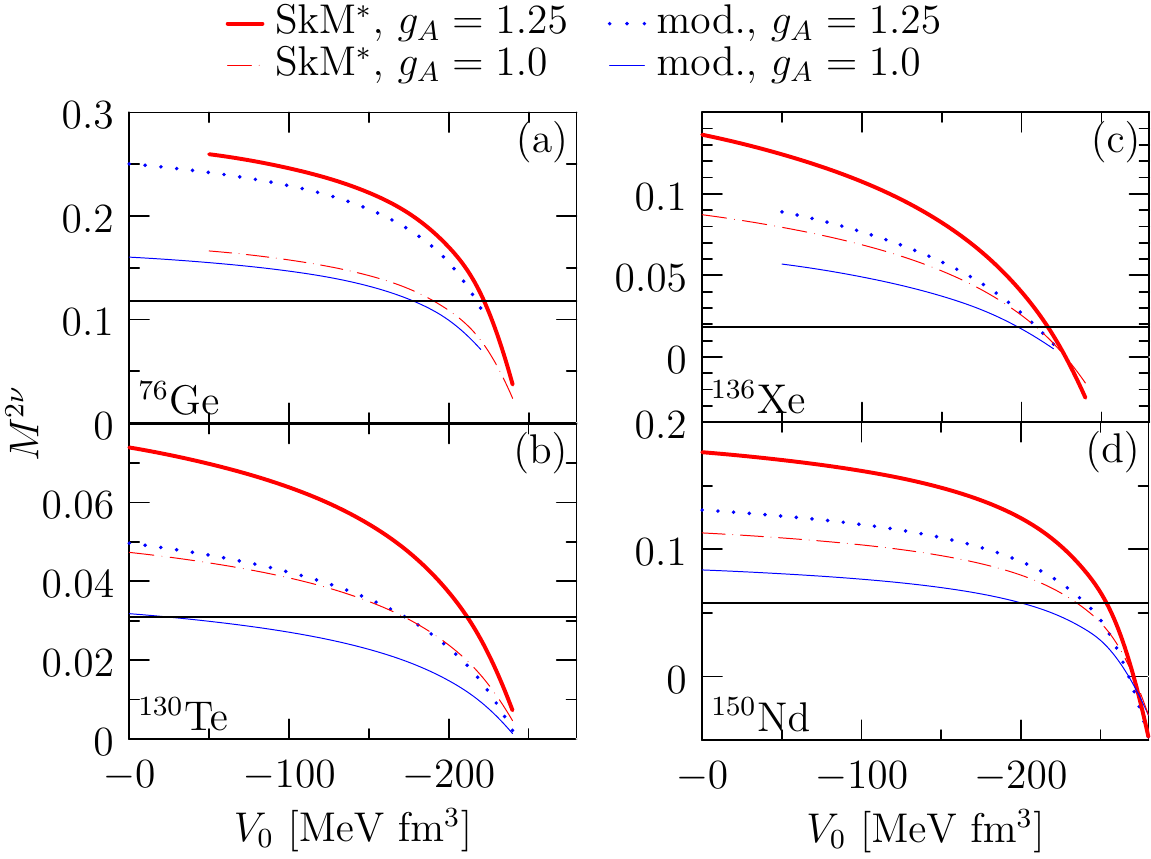}
\caption{(Color online.) The dependence of two-neutrino double-beta decay
matrix elements on $V_0$, the isoscalar pairing strength.  The thick solid and
dashed (red) curves are produced by the original SkM$^*$ interaction and the
dotted and thin (blue) curves by the modified interaction.  The thick solid and
dotted curves are computed with $g_A=1.25$, the dashed and thin solid curves
with the quenched value $g_A=1.0$.}
\label{fig:2nu}
\end{figure}

\begin{figure}[!b]
\centering
\includegraphics[width=.7\columnwidth]{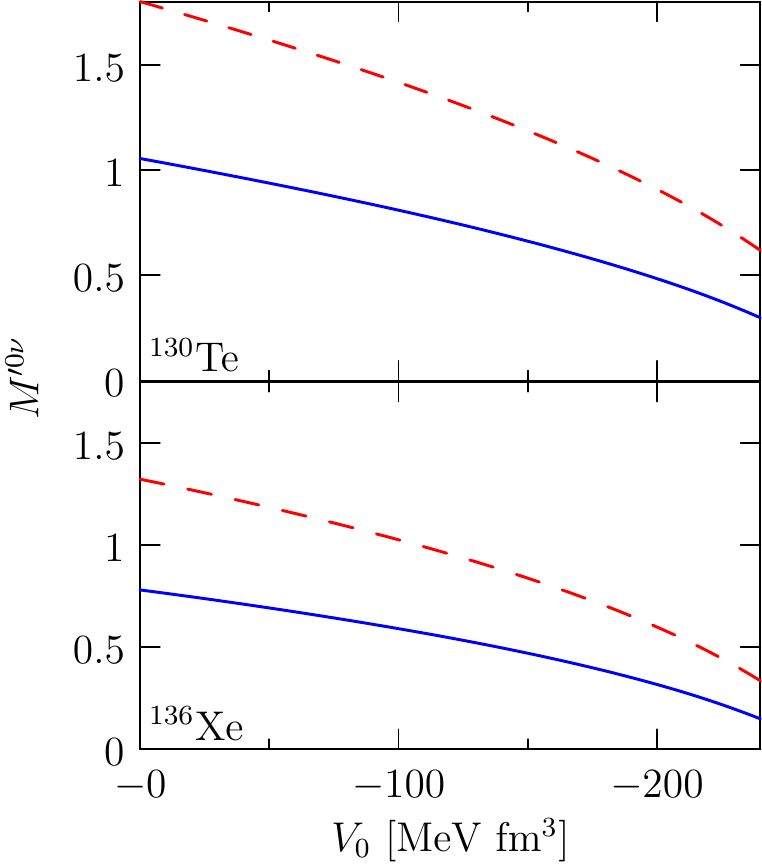}
\caption{(Color online.) The dependence of $M'^{0\nu}$ in $^{136}$Xe and
$^{130}$Te on the $V_0$ produced by the unmodified SkM$^*$ interaction.  The
solid curve represents the results with $g_A=1.0$ and the dashed curve
represents the results with $g_A=1.25$.}
\label{fig:0nu}
\end{figure}

We turn now to the matrix elements themselves.  Figure \ref{fig:2nu} displays
the dependence of the $2\nu\beta\beta$ matrix element on the isoscalar pairing
strength $V_0$ in the four systems we study.  We use the recent evaluation of
the phase-space factors in Ref.~\cite{Kot12} to extract the experimental matrix
elements.  Because $M^{2\nu}$ for $^{136}$Xe was just measured for the first
time by the EXO-200 \cite{Ack11} and KamLAND-Zen \cite{Gan12} experiments, ours
is the first QRPA double beta decay computation to use an experimentally
obtained value rather than an upper limit to determine the strength of
isoscalar pairing.

\begin{figure}[t]
\centering
\includegraphics[scale=.8]{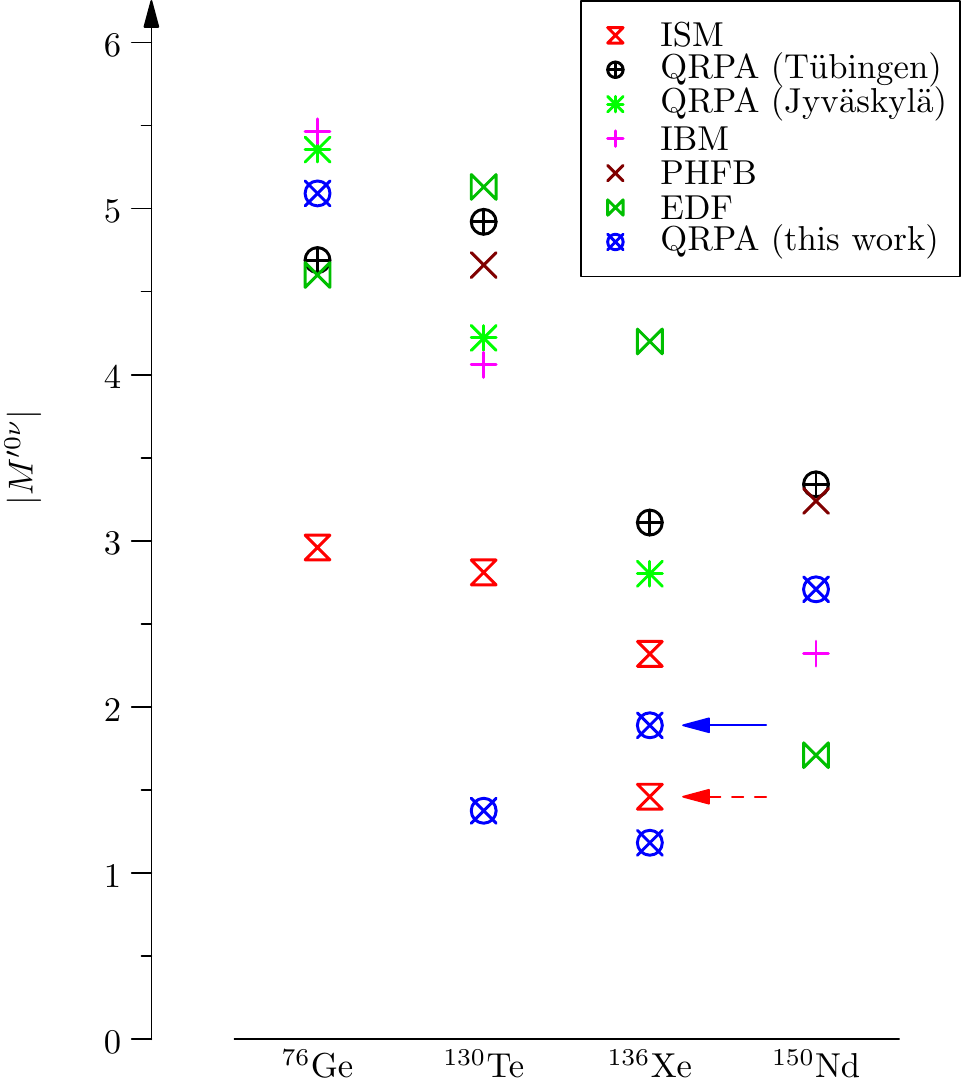}   
\caption{(Color online.) The results of Table \ref{tab:comparison} for
$g_A=1.25$.  The solid (blue) arrow points to our result with the
Hartree-Fock + Lipkin-Nogami
overlap in $^{136}$Xe. The dashed (red) arrow points to the new shell-model
result of Ref.\ \cite{Hor13} in the same nucleus.}
\label{fig:comp}
\end{figure}

Figure~\ref{fig:0nu} illustrates the dependence of the $0\nu\beta\beta$ decay
matrix element on $V_0$.  The neutrinoless matrix element is less sensitive to
this pairing mode than the two-neutrino matrix element.  We collect our final
results for the $0\nu\beta\beta$ matrix elements with both $g_A=1.25$ and
$g_A=1.0$ in Table~\ref{tab:results}.  The modification of SkM$^*$ usually
suppresses the $0\nu\beta\beta$ matrix element, by up to 15\%.  It actually
seems to increase the matrix element in $^{130}$Te by $17\%$ for $g_A=1.0$, but
as Fig.~\ref{fig:2nu} shows, the fitting procedure for $V_0$ with $g_A=1.0$
gives an anomalously small value, and so that result must be taken with a grain
of salt.  In Table~\ref{tab:comparison} we compare our values with the modified
SkM$^*$ and $g_A=1.25$ with earlier theoretical results.  Our matrix elements
for $^{76}$Ge and $^{150}$Nd are in a good agreement with the spherical result
for Ge and the deformed one for Nd in Ref.~\cite{Fan11}.  For $^{136}$Xe and
$^{130}$Te we get noticeably smaller matrix elements than obtained in prior
work, all of which was carried out in the spherical QRPA.  Figure
\ref{fig:comp} displays the same information as the table graphically. 

\begin{table}[!t]
\begin{tabular}{l c c| c c}
\hline\hline
& \multicolumn{2}{c}{SkM$^*$} & \multicolumn{2}{c}{modified SkM$^*$}   \\
\cline{2-5} & $g_A=1.0$ & $g_A=1.25$ & $g_A=1.0$ & $g_A=1.25$ \\
\hline
$^{76}$Ge    & 4.40 & 5.53 & 4.12 & 5.09  \\
$^{130}$Te & 1.13 & 1.38 & 1.32 & 1.37   \\
$^{136}$Xe & 1.26 & 1.68 & 1.18 & 1.55  \\
$^{136}$Xe (HFLN) & 1.54 & 2.05 & 1.44 & 1.89  \\
$^{150}$Nd & 2.52 & 3.14 & 2.14 & 2.71  \\
\hline\hline
\end{tabular}
\caption{The $0\nu\beta\beta$ matrix elements in our Skyrme-HFB-QRPA
calculation, with both the functional SkM$^*$ and a modified version of it, and
with both a quenched and unquenched axial-vector coupling constant $g_A$.  The
second row of $^{136}$Xe numbers contains the results with the Hartree-Fock +
Lipkin-Nogami overlap (see text). } 
\label{tab:results}
\end{table}

\begin{table}[!b]
\begin{tabular}{l c c c c c c c}
\hline\hline
& present& QRPA/T & QRPA/J & ISM & IBM-2 & PHFB & EDF \\
\hline
$^{76}$Ge  & 5.09 & 5.30, 4.69$^*$ & 5.355 & 2.96 & 5.465 &  --- & 4.60 \\
$^{130}$Te & 1.37 & 4.92           & 4.221 & 2.81 & 4.059 & 4.66 & 5.13 \\
$^{136}$Xe & 1.89 & 3.11           & 2.802 & 2.32 & 2.220 &  --- & 4.20 \\
$^{150}$Nd & 2.71 & 3.34$^*$       & ---   &  --- & 2.321 & 3.24 & 1.71 \\
\hline\hline
\end{tabular}
\caption{Comparison of our $0\nu\beta\beta$ matrix elements, from the modified
SkM$^*$ functional and $g_A=1.25$, with those obtained from the interacting
shell model (ISM, \cite{Men09}), QRPA calculations by the T\"ubingen group
\cite{sim09,Fan11} (QRPA/T) and Jyv\"askyl\"a group \cite{Suh08} (QRPA/J), the
energy-density-functional method \cite{Rod10} (EDF), projected HFB \cite{Rat10}
(PHFB) and the interacting boson model \cite{Bar09} (IBM-2).  Prior results
that include deformation are indicated by a star.  In $^{136}$Xe we use the
Hartree-Fock + Lipkin-Nogami overlap to scale our matrix element.}
\label{tab:comparison}
\end{table}

The suppression we see in $^{130}$Te can be attributed to the
deformation of the daughter nucleus.  Previous QRPA calculations for $^{130}$Te
\cite{Suh08,sim09} have assumed spherical symmetry.  We've already mentioned,
however, that a single minimum may not be adequate to represent the ground
state of $^{130}$Xe.  We suspect that the complete neglect of deformation in
previous work leads to a matrix element that is too large, but it may also be
that our sharp prolate Xe ground state yields one that is too small.  

The other decay in which we disagree significantly with previous QRPA
calculations is that of $^{136}$Xe.  Our significantly smaller result here is
not caused by deformation difference, nor does it come from the availability of
new two-neutrino decay data.  Instead, it can be traced to the overlap between
the initial and final HFB mean fields.  This overlap usually reflects the
difference in deformation between the mother and daughter nuclei, and for that
reason has been completely neglected in previous QRPA calculations for the
decay of $^{136}$Xe, where both the initial and final nuclei are spherical.  We
find here, however, that differences in pairing structure in the neutron mean
fields lead to a small overlap: $\langle \textrm{HFB}_f|\textrm{HFB}_i \rangle
= 0.47$.  The suppression is related to the $N=82$ shell closure, which
produces a sharp Fermi surface that smooths measurably with the addition of two
neutrons.  We see no reason to completely neglect the overlap, but the
situation may be analogous to the that in the decay of $^{130}$Te.  A more
realistic representation of pairing than is offered by the HFB mean field might
make the difference in structure between the initial and final nuclei a little
less dramatic.

To test that last assumption, we reevaluated the overlap in the Hartree-Fock +
Lipkin-Nogami approximation \cite{Lip60}, which reduces fluctuations in
particle number and prevents the total breakdown of pairing at closed shells.
The new occupation numbers increase the overlap to 0.57.  If we scale the
matrix element with $g_A=1.25$ and the modified Skyrme functional accordingly,
we obtain the result 1.89, indicated by the arrow in Fig.\ \ref{fig:comp}.
Although the Lipkin-Nogami calculation is not self-consistent --- we have only
modified the overlap, not the HFB quasiparticle and the QRPA strength
calculations --- this larger number is our best estimate of the matrix
element.  The deviation of the Lipkin-Nogami overlap from unity, while less
than that of the HFB overlap, still makes the result smaller than those of any
other QRPA calculations.  Interestingly, a recent shell-model \cite{Hor13}
calculation finds that increasing the model-space size produces the smallest
matrix element yet for this decay: 1.46.

All the substantial differences between our QRPA calculations and others can be
traced to deformation or pairing effects that were neglected in previous work.
Our use of a self-consistent QRPA with all nucleons treated as active
participants, the continuum accounted for, etc., doesn't, in itself, change
results dramatically.  That finding is not altogether surprising.
Self-consistency is important in the QRPA partly because it eliminates spurious
strength.  In the charge changing QRPA, however, the absence of proton-neutron
mixing in the HFB and the explicit breaking of isospin mean that there is no
spurious strength even in non-self-consistent calculations.  More importantly,
it is already well known \cite{Rod03} that differences between variants of the
QRPA largely disappear when the strength of the isoscalar pairing interaction
is adjusted so that each variant reproduces the measured two-neutrino rate.
Our variant does not escape this fate; that one parameter is a like a broad and
coarse brush that paints over any sophistication in the underlying method. 

\section{Conclusions}\label{sec:conc}

We have performed large-scale Skyrme-HFB-QRPA computations for four important
double-beta emitters.  We have allowed for axial deformation of the initial and
final nuclei.  Our implementation increases the scale of the computation to the
limits of contemporary technology. 

For $^{76}$Ge and the very deformed $^{150}$Nd, our results are in line with
the earlier results of Ref.~\cite{Fan11}.  We note, however, that the
assumption both here and elsewhere that the $^{76}$Ge and $^{76}$Se ground
states are spherical probably results in a matrix element that is too large.

In $^{130}$Te we improved on the ground state used in previous QRPA
calculations by taking into account the deformation of the final nucleus.
Shape coexistence in the daughter $^{130}$Xe is beyond the scope of the QRPA,
however; if present, it could further modify the value of the matrix element.

Our $^{136}$Xe matrix element is the first QRPA result obtained from the new
two-neutrino-decay measurements.  It is also the first to take into account the
overlap of the two sets of QRPA intermediate states.  The overlap is smaller
than one might expect because of the sharp neutron Fermi surface in the initial
nucleus.  In reality, the Fermi surface cannot be perfectly sharp, and the true
matrix element is probably better represented by replacing the HFB overlap with
the Hartree-Fock + Lipkin-Nogami overlap. The modified result is larger than
our original matrix element but still smaller than those of other QRPA
calculations.

Our computation demonstrates that there is little to be gained by further
increasing the size and sophistication of QRPA calculations.  Any
straightforward alterations to the QRPA, other than the development of a better
energy-density functional, are unlikely to improve the results substantially.
We have reached the point at which shortcomings of the QRPA itself restrict
improvement.  The inability to treat shape coexistence is an issue at least for
the daughter nuclei $^{76}$Se and $^{130}$Xe.  The mean-field treatment of
pairing may be a problem in nuclei such as $^{136}$Xe that have closed shells.
We can address these issues only by moving beyond the QRPA.

\begin{acknowledgments}
Support for this work was provided through the Scientific Discovery through
Advanced Computing (SciDAC) program funded by U.S.\ Department of Energy,
Office of Science, Advanced Scientific Computing Research and Nuclear Physics,
under award number DE-SC0008641, and by the UNEDF SciDAC Collaboration under
DOE grant DE-FC02-07ER41457.  One of us (M.T.M.) gratefully acknowledges
fruitful discussions with Prof.\ Mihai Horoi, alongside the support and
hospitality he extended at Central Michigan University during the spring of
2012. 

This work used the Extreme Science and Engineering Discovery Environment
(XSEDE), which is supported by National Science Foundation grant number
OCI-1053575.  Most of the computations were performed on Kraken at the National
Institute for Computational Sciences (http://www.nics.tennessee.edu/).  This
research also used resources of the National Energy Research Scientific
Computing Center, which is supported by the Office of Science of the U.S.
Department of Energy under Contract No.  DE-AC02-05CH11231.
\end{acknowledgments}

\appendix*

\section{Neutrinoless double beta decay matrix elements in the cylindrical box}\label{sec:appendix}

Equation \eqref{monu-detail} is in essence a trace of a product of four large
square matrices.  The transition densities in that equation are
\begin{equation}
	\langle 0^+_f | c_{-p}^\dag c_n | N \rangle = s_p v_p u_n X_{pn}^N +
	u_p s_n v_n Y_{-p-n}^N \,,
\end{equation}
and
\begin{equation}
	\langle N' | c_{p'}^\dag c_{-n'} | 0^+_i \rangle = -u_{p'} s_{n'}
	v_{n'} X_{p'n'}^{N'} - s_{p'} v_{p'} u_{n'} Y_{p'n'}^{N'} \,.
\end{equation}
Here the indices $p$ and $p'$ indicate protons, and $n$ and $n'$ neutrons, as
discussed in the main text; $c_p^\dag$ is a proton creation operator, and $u_p$
and $s_p v_p$ are the proton occupation amplitudes in the canonical basis, in
the notation of Ref.\ \cite{ring-schuck}.  $X_{pn}^N$ and $Y_{-p-n}^N$ are
forward-going and backward-going QRPA amplitudes.

To reduce the number of nested numerical integrals in the $0\nu\beta\beta$
matrix elements in Eq.\ \eqref{monu-detail}, we take advantage of the following
expansion for the spherical Bessel function in Eq.\ \eqref{eq:monusp}:
\begin{equation} 
j_0(qr_{ab}) = 4\pi \sum_{l=0}^\infty
j_l(qr_a) j_l(qr_b) \sum_{m=-l}^l Y_{lm}^* (\hat r_a) Y_{lm} (\hat r_b) \,.
\end{equation}
This allows us to separate the integrals over coordinates of the two nucleons:
\begin{align}
	K^{F}_{pn,p'n'} =& \int_0^\infty dq \: \frac{qh_\mathrm{F} (q^2)}{q +
	E_\mathrm{ave}} \sum_{l=K}^\infty (2l+1) \frac{(l-K)!}{(l+K)!} 
	\nonumber \\
	&\times I_{-pn}^{lK} (q) I_{p'-n'}^{lK} (q) \,,
\end{align}
and
\begin{align}
	K^{GT}_{pn,p'n'} =& \int_0^\infty dq \: \frac{qh_\mathrm{GT} (q^2)}{q + E_\mathrm{ave}}
	\sum_{\mu=-1}^{1} (-1)^\mu \nonumber \\
	&\times \sum_{l=\mathrm{max}(0,K-\mu)}^\infty (2l+1)
	\frac{(l-(K-\mu))!}{(l+(K-\mu))!} \nonumber\\ 
	&\times I_{-pn}^{l,K-\mu,-\mu} (q) I_{p'-n'}^{l,K-\mu,\mu} (q) \, ,
\end{align}
where $E_\mathrm{ave} = \bar E -(E_i + E_f)/2$.  Naturally, the infinite
summations over $l$ must be truncated.  For most values of the neutrino energy
$q$, not many terms are needed for convergence.  In the program we truncate the
expansion dynamically by requiring a preset accuracy in the quadrature for each
value of $q$.

The axial symmetry of the normalized canonical single-particle wave functions
means that they can be written in the form 
\begin{equation}
	\Psi_a(\vec r) = \frac{1}{\sqrt{2\pi}} \sum_{s = \pm 1/2} \psi_a
	(s;\rho,z) e^{i(j_a^z - s)\phi} \chi_s \,,
\end{equation}
where $s$ is the spin projection, $\chi_s$ is a standard two-component spinor,
and $j^z_a$ the angular-momentum projection onto the intrinsic axis.  The
integrations over the azimuthal angle $\phi$ are trivial and the integrals
$I_{ab}^{lm} (q)$ and $I_{ab}^{lm\nu} (q)$ are therefore only two-dimensional:
\begin{equation} 
\begin{split} 
	I_{ab}^{lm} (q) =& \int_{-\infty}^\infty dz
\int_0^\infty d\rho \: \rho \: \psi_a^\dag(\rho,z) \psi_b(\rho,z) \\ &\times
j_l (q \sqrt{\rho^2 + z^2}) P_l^m \left( \frac{z}{\sqrt{\rho^2 + z^2}} \right)
\,,
\end{split} 
\end{equation}
and
\begin{equation}
	\begin{split}
	I_{ab}^{lm\nu} (q) =& \int_{-\infty}^\infty dz \int_0^\infty d\rho \: \rho \: \psi_a^\dag(\rho,z) \sigma_\nu \psi_b(\rho,z) \\ &\times
	j_l (q \sqrt{\rho^2 + z^2}) P_l^m \left( \frac{z}{\sqrt{\rho^2 + z^2}}
	\right) \,.
	\end{split}
\end{equation}
Here the $P_l^m (x)$ are the usual associated Legendre polynomials,
$\sigma_\nu$ are the Pauli matrices in the spherical vector basis and 
\begin{equation}
	\psi_a(\rho, z) = \begin{pmatrix}\psi_a(+1/2; \rho,z) \\ \psi_a(-1/2;
	\rho,z)\end{pmatrix} \,.
\end{equation}

As discussed in the main body of the text, we also need to evaluate the
overlaps $\langle N | N' \rangle$ between the QRPA states stemming from
different mean fields.  A satisfactory expression for these is lacking, but the
chief ingredient in any such expression will be the overlap of two HFB vacua.
The generalized Thouless theorem \cite{ring-schuck} relating the two
non-orthogonal quasiparticle vacua $|\textrm{HFB}_i\rangle$ (initial state) and
$|\textrm{HFB}_f\rangle$ (final state) to each other is:
\begin{equation}\label{eq:vacua}
	| \textrm{HFB}_i \rangle = \mathcal{N}^{-1} \exp \left( \sum_{kl}
	D_{kl} a_k^{(f)\dag} a_l^{(f) \dag} \right) | \textrm{HFB}_f \rangle \,,
\end{equation}
where the $a_k^{(f)\dag}$ are quasiparticle creation operators in the final
nucleus.  The normalization factor is related to the transformation
coefficients $D_{kl}$ via the Onishi formula:
\begin{equation}
	\mathcal{N} = \langle \textrm{HFB}_f | \textrm{HFB}_i \rangle^{-1} =
	\sqrt{\det (1 + D^\dag D)} \,.
\end{equation}
Because the canonical-basis wave functions form a complete set, there exists a
linear transformation between the two HFB solutions:
\begin{equation}\label{eq:qptrans}
	a_k^{(f)\dag} = \sum_{n} (\mathcal{R}_{kn} a_n^{(i)\dag} +
	\mathcal{S}_{k,-n} a_{-n}^{(i)}) \,,
\end{equation}
where
\begin{equation}
	\mathcal{R}_{kn} = \langle n | k \rangle (u_k u_n + s_k v_k s_n v_n) \,,
\end{equation}
and 
\begin{equation}
	 \mathcal{S}_{k,-n} =  \langle n | k \rangle (u_k s_n v_n - s_k v_k
	 u_n) \,.
\end{equation}
Substituting Eq.\ \eqref{eq:vacua} and \eqref{eq:qptrans} into the definition of the quasiparticle vacuum
\begin{equation}
	a_{-k}^{(f)} | \textrm{HFB}_f \rangle = 0 \,,
\end{equation}
expanding the exponential, and comparing the terms containing one quasiparticle
creation operator, we get the matrix equation
\begin{equation}
	\mathcal{R}^* D = -\mathcal{S}^* \,,
\end{equation}
from which we can obtain the transformation coefficients $D_{kl}$. 

As mentioned earlier, we approximate the QRPA overlaps states by QTDA overlaps,
i.e.\ by neglecting the $Y$'s.  This leads finally to the expression
\begin{widetext}
\begin{equation}
	\langle N | N' \rangle = \mathcal{N}^{-1} \sum_{pn} \sum_{p'n'}
	X_{pn}^{N*} X_{p'n'}^{N'} \left(\mathcal{R}_{p'p} + \sum_{p''}
	\mathcal{S}_{p'p''} D_{p''p}
	\right) \left(\mathcal{R}_{n'n} + \sum_{n''} \mathcal{S}_{n'n''}
	D_{n''n} \right) \,.
\end{equation}
\end{widetext}
This formula differs slightly from the one presented in Ref.\ \cite{Sim04} and
used in most QRPA double-beta-decay calculations.  Our overlap differs in that
we keep the transformation between the two HFB bases accurate and neglect the
usually tiny term proportional to two $Y$ amplitudes.  In test calculations we
find the numerical difference between the two prescriptions to be negligible,
as the common leading term is already a good approximation.  A more consistent
evaluation of these overlaps that includes ground-state correlations can easily
get both very complicated and computationally demanding, as evidenced by recent
work in the like-particle QRPA in Ref.\ \cite{Ter12}.

\end{document}